\documentclass[aps,pra,reprint,superscriptaddress]{revtex4-1}
\usepackage{graphicx}
\usepackage{amsmath}
\usepackage{amssymb}
\usepackage[colorlinks,linkcolor=blue,
            citecolor=blue,
            hyperindex,
            pdfstartview=FitH,
            plainpages=false]
            {hyperref}
\usepackage{soul}

\newcommand {\bscco}{Bi$_2$Sr$_2$CaCu$_2$O$_{8+\delta}$}

\newcommand {\uJcm}{$\mu$J/cm$^{2}$}
\newcommand {\PDH}{PDHS} 

\begin{document}
\title{Interplay of superconductivity and bosonic coupling in the peak-dip-hump structure of \bscco{}}

\author{Tristan L. Miller}
\affiliation{Materials Sciences Division, Lawrence Berkeley National Laboratory, Berkeley, California 94720, USA}
\affiliation{Department of Physics, University of California, Berkeley, California 94720, USA}
\author{Wentao Zhang}
\affiliation{Department of Physics and Astronomy, Shanghai Jiao Tong University, Shanghai 200240, China.}
\author{Jonathan Ma}
\affiliation{Materials Sciences Division, Lawrence Berkeley National Laboratory, Berkeley, California 94720, USA}
\affiliation{Department of Physics, University of California, Berkeley, California 94720, USA}
\author{Hiroshi Eisaki}
\affiliation{Electronics and Photonics Research Institute, National Institute of Advanced Industrial Science and Technology, Ibaraki 305-8568, Japan}
\author{Joel E. Moore}
\author{Alessandra Lanzara}
\email{alanzara@lbl.gov}
\affiliation{Materials Sciences Division, Lawrence Berkeley National Laboratory, Berkeley, California 94720, USA}
\affiliation{Department of Physics, University of California, Berkeley, California 94720, USA}
\date {\today}

\begin{abstract}
Because of the important role of electron-boson interactions in conventional superconductivity, it has long been asked whether any similar mechanism is at play in high-temperature cuprate superconductors.  Evidence for strong electron-boson coupling is observed  in cuprates with angle-resolved photoemission spectroscopy (ARPES), in the form of a dispersion kink and peak-dip-hump structure.  What is missing is evidence of a causal relation to superconductivity.  Here we revisit the problem using the technique of time-resolved ARPES on \bscco{}.  We focus on the peak-dip-hump structure, and show that laser pulses shift spectral weight into the dip as superconductivity is destroyed on picosecond time scales.  We compare our results to simulations of Eliashberg theory in a superconductor with an Einstein boson, and find that the magnitude of the shift in spectral weight depends on the degree to which the bosonic mode contributes to superconductivity.  Further study could address one of the longstanding mysteries of high-temperature superconductivity.

\end{abstract}

\maketitle

\section{Introduction}

In conventional superconductors, the electron-phonon coupling is the ``glue" that allows Cooper pairing, and leads to superconductivity \cite{Bardeen1957}.  In high-temperature cuprate superconductors, electrons form Cooper pairs, and have been found to strongly couple to a bosonic mode, but it is unclear whether these two facts are related to each other.  Critics observe that cuprates are dominated by the Hubbard repulsion and antiferromagnetic exchange coupling, and that any bosonic interaction appears to be of lesser importance \cite{Anderson2007}.  Experiments have extensively explored the electron-boson interaction, but have only begun to answer this fundamental question \cite{Bok2016}.

Early evidence for electron-boson coupling in cuprates was found with angle-resolved photoemission spectroscopy (ARPES), in the form of a peak-dip-hump structure (\PDH{}) in the energy profiles of ARPES intensity \cite{Dessau1991, Hwu1991, Dessau1992, Fedorov1999}.  The \PDH{} was observed most strongly at the antinodal point below $T_c$.  The \PDH{} was eventually understood as one aspect of an electron dispersion kink \cite{Norman1997, Campuzano1999, Kaminski2001, Lanzara2001, Gromko2003, Kim2003, Sato2003, Cuk2004, Plumb2013}.  In this picture, the dispersion kink is a sudden drop in quasiparticle lifetime, arising from the decay pathway by boson emission.  A quasiparticle with a long lifetime is observed as a sharp peak in intensity, while a quasiparticle with short lifetime is observed as a broad hump.  The \PDH{} appears when the peak and hump are seen together in ARPES energy profiles.

Other experimental techniques, such as scanning tunneling microscopy \cite{Fischer2007} and Raman spectroscopy \cite{Loret2016}, have observed features that are similar to the \PDH{}.  However, in this context the \PDH{} does not have exactly the same interpretation. The difference is that ARPES studies look at a particular point in momentum space, while these other techniques effectively integrate over a large range of momentum.  At a single momentum, the peak represents a single quasiparticle pole; but in a momentum-integrated measurement, the peak represents a pileup of many quasiparticle poles at the edge of the superconducting gap.  Some simulations have shown that electron-boson coupling could create a momentum-integrated \PDH{} \cite{Zasadzinski2003}, but other studies have argued that it arises from charge order or the pseudogap \cite{Gabovich2014, Loret2016}.  Several recent studies have demonstrated the value of integrating across just one dimension of momentum \cite{Reber2012, Reber2013, Parham2013, Miller2017}.  This analysis sacrifices resolution perpendicular to the Fermi surface, but maintains resolution along the Fermi surface and greatly improves statistics.  

Time- and angle-resolved photoemission spectroscopy (TARPES) \cite{Smallwood2012RSI,Ishida2014,Smallwood2016EPL} is another technique that could address the electron-boson coupling and its link to superconductivity.  TARPES uses a short laser pulse to pump a material out of equilibrium, and after a given time delay, uses a second laser pulse to take an ARPES measurement.  In the cuprate \bscco{} (Bi2212), pump pulses with fluence on the order of 15 \uJcm{} can partially or completely suppress superconductivity on a picosecond time scale, leaving a transient pseudogap state \cite{Smallwood2012, Smallwood2014, Ishida2016, Zhang2017, Boschini2018}.    During this time, about 1 ps after pumping, electrons are in quasithermal equilibrium, although they are not yet in thermal equilibrium with other degrees of freedom in the system \cite{Perfetti2007, Graf2008}.  So far, TARPES studies have addressed electron-boson coupling by investigating its impact on quasiparticle dynamics \cite{Perfetti2007, Smallwood2012, Smallwood2015, Yang2015, Rameau2016}, or by focusing on the electron-boson kink along the superconducting node \cite{Zhang2014,Ishida2016,Rameau2014}.

Here we look at the \PDH{} in Bi2212, applying TARPES and integrating along one-dimensional momentum cuts.  Laser pulses cause spectral weight to shift into the dip, causing an increase of intensity near 70 meV (near the kink).  We quantify the strength of the dip by measuring the magnitude of the increase of intensity, finding that it is stronger away from the node, and roughly equal in underdoped and overdoped samples. We also observe an increase of intensity near 140 meV, which could indicate a second kink or a second-order effect of the first kink.

To understand the \PDH{} in the context of one-dimensional momentum integration, we built a simulation using Eliashberg theory in a superconductor with an Einstein boson.  We find that a \PDH{} appears only when both superconductivity and electron-boson coupling are simultaneously present.  Furthermore, the strength of the dip depends on the extent to which the boson couples to the $d$-wave superconducting parameter.  Further study may be able to determine whether and how much the bosonic mode is involved in the mechanism of high-temperature superconductivity.

\section{Materials and Methods}

We use the TARPES setup described in Ref. \cite{Smallwood2012RSI}.  Samples of Bi2212 were cleaved \em{}in situ\em{} at base pressures less than $5 \times 10^{-11}$ Torr.  An infrared (1.48 eV) laser pulse with fluence of 24 \uJcm{} was used to pump samples.  Electrons were ejected with an ultraviolet (5.93 eV) pulse, to be measured by a SPECS Phoibos 150 mm hemispherical analyzer. The time resolution is ~ 300 fs, the energy resolution is ~ 23 meV, and the momentum resolution is ~ 0.003 $\mathrm{\AA}^{-1}$.

Because of the precision required for this study, we correct for several known systematic errors.  We correct for the nonlinear sensitivity of our detector \cite{Smallwood2012RSI}.  All data were taken in a fixed mode, which can create problems because some parts of the camera are more sensitive than others, and we correct for this by taking data accumulated over several days and measuring the systematic bias in intensity.  We partially correct for the instrumental energy resolution by applying five iterations of the Lucy-Richardson deconvolution algorithm \cite{Lucy1974}.  Over the course of a measurement, laser power may drift, which causes the Fermi level to drift due to space charging \cite{Zhou2005}.  Therefore, we take all measurements in cycles, and correct for any drift in the Fermi level (never more than 0.2 meV per cycle).  Finally, laser pumping induces a shift in chemical potential on the order of 2 meV, and while this has a physical origin \cite{Miller2015,Miller2017}, we correct for the shift as it is not the focus of this study.

\section{Pump-induced intensity difference}

\begin{figure}\centering\includegraphics[width=3.4in]{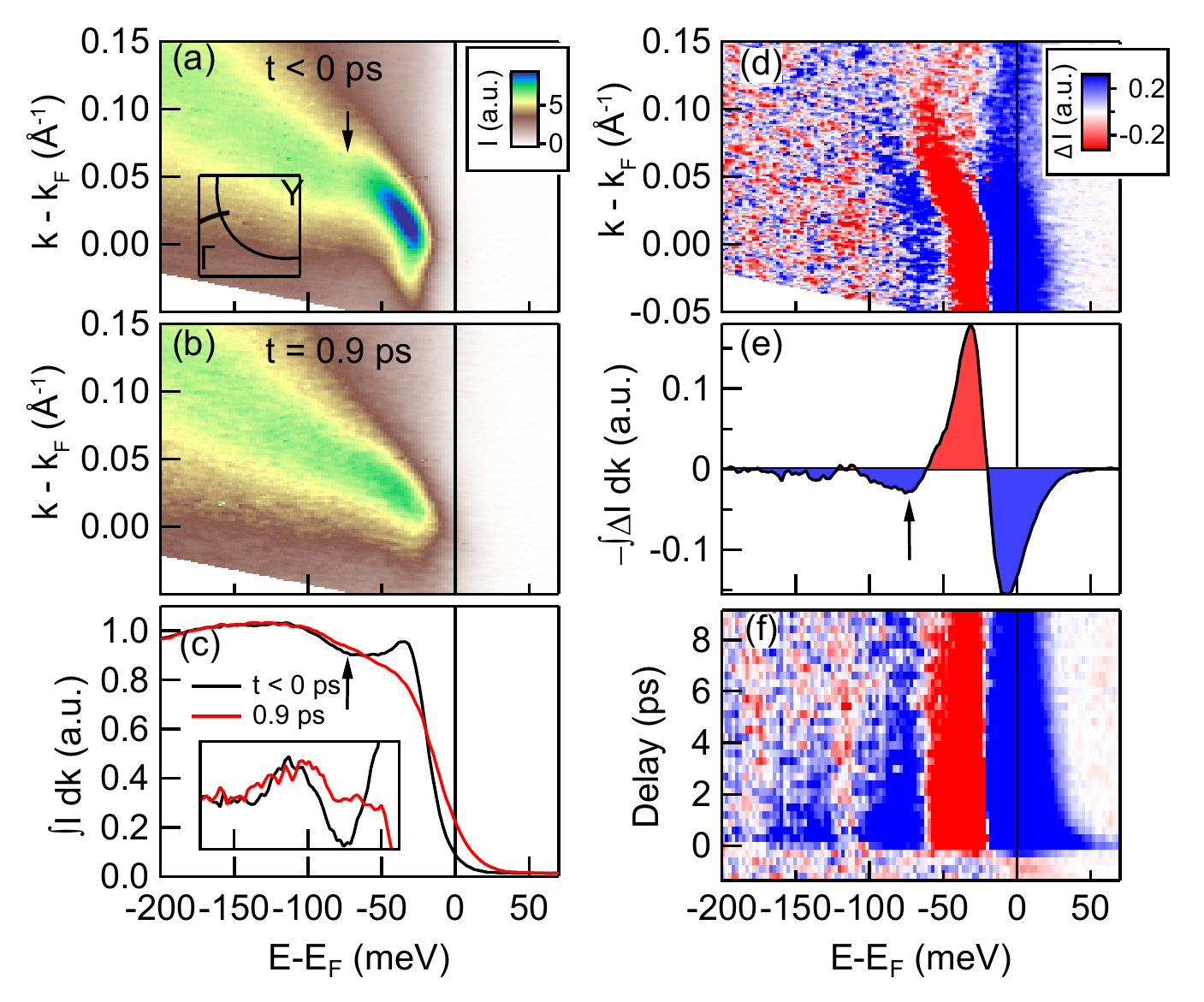}
\caption{(a) The ARPES intensity map of an overdoped Bi2212 sample ($T_c$ = 70 K) at equilibrium at 20 K.  The inset shows where in momentum space the cut was taken.  (b) A map of the same sample 0.9 ps after pumping with a 24 \uJcm{} pulse.  (c) The intensity in (a) and (b) integrated over momentum.  The inset shows a closeup of the two curves (after subtracting a quadratic fit).  (d) A map of the change in intensity between (a) and (b).  (e) The momentum-integrated change of intensity at $t$ = 0.9 ps.  (f) The momentum-integrated change of intensity plotted as a function of delay time.  
}
\label{Fig1}
\end{figure}

To introduce our technique, we begin with an overdoped (OD) Bi2212 sample ($T_c$ = 70 K), which is shown in Fig. \ref{Fig1}.  Fig. \ref{Fig1}(a) shows the ARPES intensity map at equilibrium at 20 K, and Fig. \ref{Fig1}(b) shows the same map 0.9 ps after the pump pulse.  We point out some basic features of the spectra that would be familiar to ARPES experts.  The arrow roughly indicates the position of the kink, a bend in the electronic dispersion that separates sharp coherent peaks at low binding energy from broad incoherent peaks at high binding energy. Note that the energy axis is customarily inverted, such that higher binding energy corresponds to more negative $E-E_F$.  Upon pumping, the kink is significantly weakened, although it does not disappear entirely.  The signature of superconductivity is a gap between the ARPES intensity and the Fermi energy $E_F$.  Although superconductivity is suppressed by pumping, a gap remains because of the transient pseudogap state \cite{Smallwood2014, Ishida2016}.  This particular measurement is taken along the momentum cut indicated in the inset of Fig. \ref{Fig1}(a), where both the superconducting gap and pseudogap are present.

Figure \ref{Fig1}(c) shows ARPES intensity integrated over momentum.  The integration window is given by our measurement window, which is slightly larger than the maps shown.  To account for fluctuations in laser power, the intensity curves are normalized by the intensity averaged at high binding energy (200 to 300 meV).  The intensity at equilibrium shows a clear \PDH{}, but 0.9 ps after pumping, the peak and dip have been suppressed.  The inset shows a closeup of the two curves, and one can see that pumping causes intensity to fill into the dip.  These observations motivated us to consider the difference between curves.

Figure \ref{Fig1}(d) shows a map of the intensity difference ($\Delta I$), and Fig. \ref{Fig1}(e) shows $-\Delta I$ integrated over momentum.  Fig. \ref{Fig1}(e) uses the same units as Fig. \ref{Fig1}(c), so that 0.1 is approximately a 10$\%$ change in the spectral weight.  The increase of intensity near $E_F$ (blue region) corresponds to intensity filling the superconducting gap, and an increase in the occupation fraction of the quasiparticle states above $E_F$.  The decrease of intensity near 30 meV (red region) corresponds to the suppression of the peak in the \PDH{}, as well as a decrease in the occupation fraction of the quasiparticle states.  The most striking observation is an increase of intensity near 70 meV (indicated by arrow) which matches with the dip in Fig. \ref{Fig1}(c) and the kink in Fig. \ref{Fig1}(a).  This increase of intensity cannot be explained by a change in the occupation fraction of electrons, and implies a change in the density of states itself.  Note that this is similar to measurements of the momentum-resolved \PDH{} at equilibrium, which find that spectral weight shifts from the peak into the dip as the temperature crosses $T_c$ \cite{Dessau1991, Hwu1991, Dessau1992, Fedorov1999, Norman1997, Campuzano1999, Kaminski2001}; however, most earlier measurements were taken at the antinode, and did not have as much power to measure the \PDH{} near the node.

In Fig. \ref{Fig1}(f), we show the delay dependence of the momentum-integrated $\Delta I$.  It is clear that $\Delta I$ appears nearly immediately after pumping, and decreases over several picoseconds, but its qualitative features do not change during that time.  The saturated color scale also make clear an increase of intensity near 140 meV, which will be discussed further later.

\section{Momentum and doping dependence}

\begin{figure*}\centering\includegraphics[width=7in]{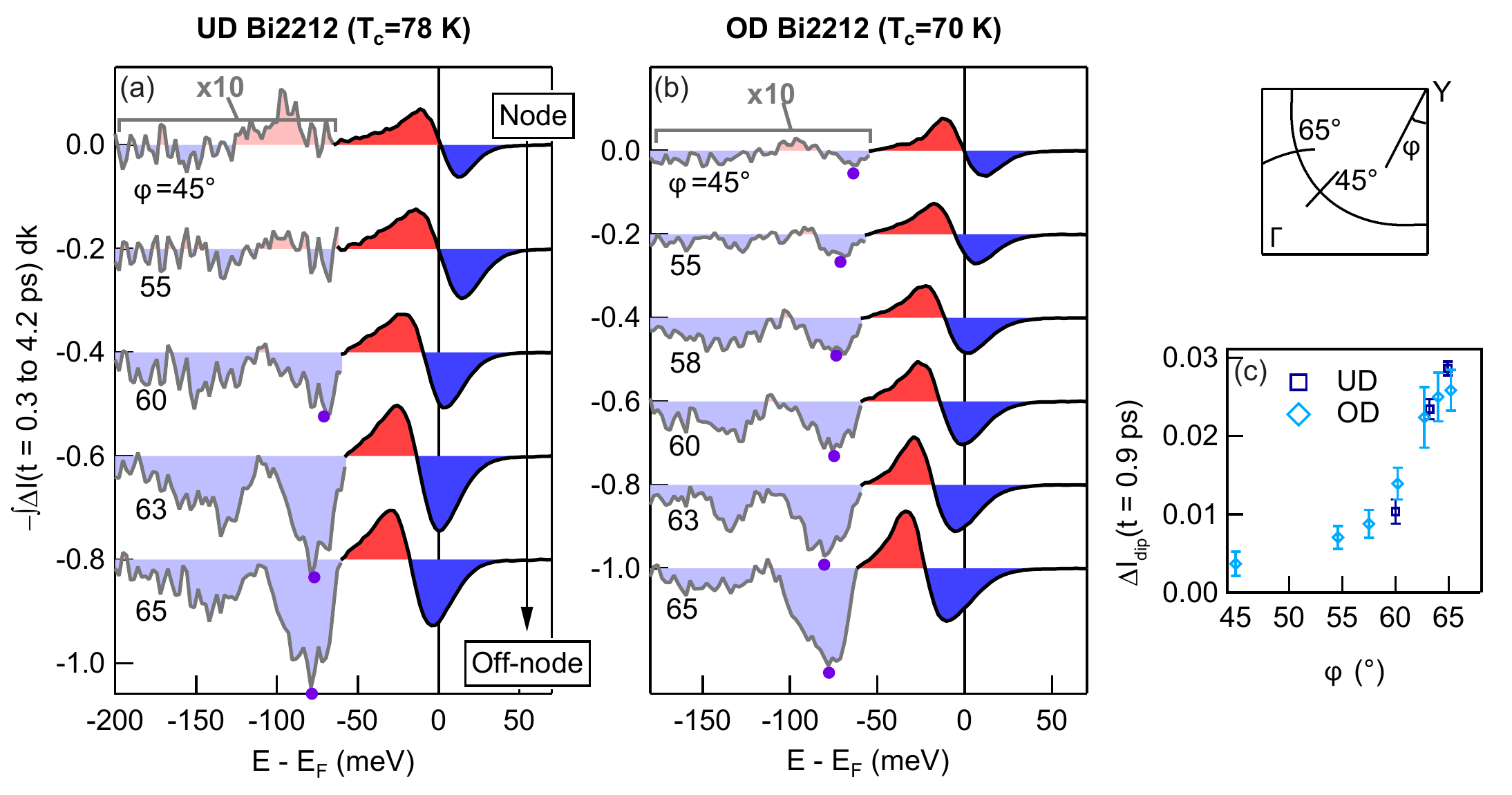}
\caption{The momentum-integrated $\Delta I$ curves for an underdoped Bi2212 sample (a) and an overdoped sample (b).  Each curve corresponds to a different momentum cut (parametrized by $\varphi$, defined in the upper right panel), and the curves are offset for clarity.  Parts of each curve have been magnified by a factor of 10.  The curves are averaged from 0.3 to 4.2 ps to improve signal to noise ratio; this does not qualitatively change the results.  Purple circles indicate minima in the curves.  (c) The change of intensity at 0.9 ps, averaged over a 20 meV window centered at the purple circles.
}
\label{Fig2}
\end{figure*}

In the previous section, we took overdoped Bi2212 and integrated $\Delta I$ over a particular momentum cut.  The next step is to expand this measurement to other momenta and other dopings.  Figures \ref{Fig2}(a) and \ref{Fig2}(b) show the $-\Delta I$ curves measured on underdoped (UD) Bi2212 ($T_c$ = 78 K), and OD Bi2212 ($T_c$ = 70 K), both measured at 20 K.   Each curve is taken at a different momentum cut, parametrized by the Fermi surface angle $\varphi$, defined in the upper right panel.  In order to compare different curves, we normalize each measurement such that the equilibrium intensity at 100 meV is 1.  In each curve, we identify minima near 70 meV, corresponding to the dip.  We define the dip strength $\Delta I_{\mathrm{dip}}$, shown in Fig. \ref{Fig2}(c), to be the change of intensity averaged in a 20 meV window centered at the minimum.


In the UD sample, a dip cannot be identified near the node; but apart from this, $\Delta I_{\mathrm{dip}}$ is similar in the UD and OD samples.  This disconfirms any relation between the dip and the pseudogap, since the pseudogap is much smaller in the overdoped sample \cite{Vishik2012}.  It also suggests that the bilayer splitting, which sometimes causes a \PDH{} in overdoped samples near the antinode \cite{Feng2001, Gromko2002, Kordyuk2002, Borisenko2003}, is not the source of the \PDH{} in our measurements.

We also observe that $\Delta I_{\mathrm{dip}}$ increases with $\varphi$.  This is consistent with previous studies, which all agree that the \PDH{} is much stronger near the anti-node \cite{Dessau1991, Hwu1991, Dessau1992, Fedorov1999, Norman1997, Campuzano1999, Kaminski2001}.  The momentum dependence of the \PDH{} is typically attributed to the electron-boson coupling strength, which, in the superconducting state, increases with distance from the node \cite{Kaminski2001, Sato2003, Gromko2003}.  However, we will later propose that the momentum dependence of $\Delta I_{\mathrm{dip}}$ is related to the size of the gap.

Finally, we note additional minima appearing near 140 meV.  Given the similarity between the minima at 70 meV and 140 meV, it is tempting to identify the feature at 140 meV with an additional dip structure.  Such additional dip structures have previously been observed in sensitive ARPES measurements \cite{He2013}.

\section{Characteristic energy scales}

\begin{figure}\centering\includegraphics[width=3.4in]{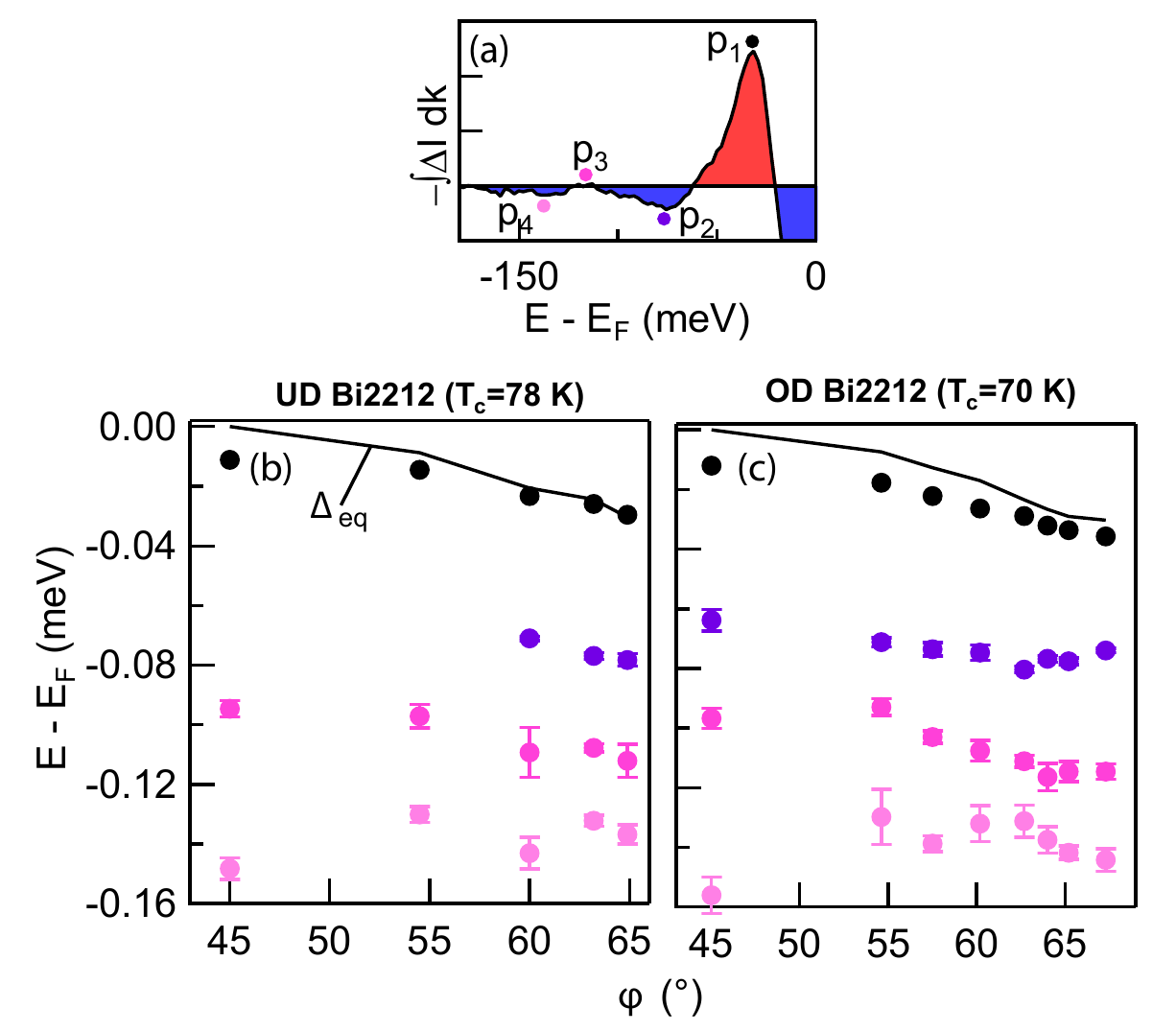}
\caption{(a) Illustration of several characteristic energy scales observed in $\int \Delta I \mathrm{d}k$.  $p_i$ is the energy of the $i$th minima or maxima below the $E_F$.  These characteristic energy scales are shown as a function of Fermi surface angle in UD Bi2212 (b) and OD Bi2212 (c).  The size of the gap at equilibrium ($\Delta_{\mathrm{eq}}$) is estimated with standard fitting methods \cite{Norman1998}. Error bars are determined by the variance between measurement cycles, although measurements may also be affected by systematic error.
}
\label{Fig3}
\end{figure}

In Fig. \ref{Fig3}, we identify local minima and maxima in the momentum integrated intensity curves, and show how these energy scales depend on doping and momentum.  Away from the node, $p_1$ corresponds to the peak in the \PDH{}, which arises from a pileup of quasiparticle states at the superconducting gap edge.  At the superconducting node, there is no superconducting gap, and $p_1$ instead corresponds to the increase in electronic temperature.  Thus, as we move away from the node, $p_1$ starts near electronic temperature scale, and ends near the superconducting gap size $\Delta_\mathrm{eq}$.

The energy scale $p_2$ corresponds to the dip in the \PDH{}, which nearly matches the kink energy.  The kink energy is significant because it is expected to appear at $\Delta+\Omega$, where $\Delta$ is the superconducting gap size and $\Omega$ is the energy of the bosonic mode \cite{Campuzano1999}.  $\Delta$ is commonly taken to be the size of the superconducting gap at the same momentum \cite{Cuk2004, Devereaux2004, Kulic2005, Plumb2013}, but is sometimes interpreted to be the maximum gap size \cite{Lee2007a, Lee2008}.  However, we note that $p_2$ is not a very precise way to estimate the kink energy, since $p_2$ is not exactly the energy of the dip, nor is the dip at exactly the same energy as the kink.  In earlier measurements of OD Bi2212 samples, the kink appears near 60 meV at the node \cite{Lanzara2001} and near 40 meV at the antinode \cite{Cuk2004}.  This does not match the energy of $p_2$ in Fig. \ref{Fig3}(c) which appears near 60 meV at the node, and nearly 80 meV away from the node.  Later, simulations will corroborate our conclusion that $p_2$ may deviate slightly from the kink energy.

The energy scales $p_3$ and $p_4$ are related to the mysterious feature at 140 meV.  One interpretation is that this feature arises from another bosonic coupling mode at energy $\Omega'$, with $p_4$ appearing near $\Delta+\Omega'$.  A second possibility is that they are a second-order feature of the first bosonic coupling mode, with $p_4$ appearing near $\Delta+2\Omega$.

\section{Simulations of momentum-integrated intensity}

The \PDH{} has a different interpretation depending on whether it is measured at a single momentum, or using momentum-integrated techniques.  Here we develop a framework to interpret the \PDH{} in ARPES spectra integrated over a single dimension of momentum, and we support this framework with a simulation.

Our framework is based on the simulation in Ref. \cite{Sandvik2004}.  We begin with the Nambu-Gor'kov formalism, which treats the electron Green's function $G$ and self-energy $\Sigma$ as a 2x2 matrices \cite{Scalapino1969}.  The diagonal components relate to single particles, while the off-diagonal components relate to electron pairs.  The Green's function is given by the Dyson equation
\begin{equation}
G^{-1}(k,\omega) = \omega \mathbf{1} - \epsilon(k)\tau_3 - \Sigma(k,\omega)\label{NambuDyson},
\end{equation}
where $k$ is momentum, $\omega$ is energy, $\tau_i$ are the Pauli matrices, and $\epsilon(k)$ is the bare electron dispersion.  The self-energy is has the canonical form
\begin{multline}
\Sigma(k,\omega) = (1-Z(k,\omega))\omega \mathbf{1} + \phi(k,\omega) \tau_1 \\
 + \bar\phi(k,\omega)\tau_2 + \chi(k,\omega) \tau_3\label{CanonicalSelfEnergy},
\end{multline}
which defines the complex-valued functions $Z$, $\phi$, $\bar\phi$ and $\chi$.  $\bar\phi$ can be neglected by choice of phase convention, and $\chi$ can be neglected when the slope of $\epsilon(k)$ is assumed to be constant.  The ARPES intensity is given by
\begin{equation}
I(k,\omega) = -(1/\pi) f(\omega) I_0(k) \mathrm{Im}~G_{11}(k,\omega) \label{ARPESIntensity},
\end{equation}
where $f(\omega)$ is the electron distribution function, $I_0(k)$ is the square of the dipole matrix element \cite{Damascelli2003}, and $G_{11}$ is the upper left entry of $G$.

To calculate $\int I(k,\omega) \mathrm{d}k$, we make two approximations.  First, we assume that the integration window is large enough to cover most of the intensity of the dispersion.  Second, we assume that $\Sigma$, $I_0$, and the bare electron velocity are independent of momentum within the window of integration.  With these approximations, the momentum integrated intensity obeys the form
\begin{equation}
f(\omega)~\mathrm{Re}\frac{Z(\omega)\omega}{\sqrt{(Z(\omega)\omega)^2-\phi(\omega)^2}}\label{DIeq}.
\end{equation}

Equation (\ref{DIeq}) generates a \PDH{} only when two conditions are simultaneously fulfilled.  First, $\phi$ must be nonzero, or else the integrated intensity reduces to $f(\omega)$, a function with no \PDH{}.  This implies the presence of a superconducting gap.  Second, either $Z$ or $\phi$ must have some dependence on energy; a kink fulfills this condition.  In our equilibrium data (except those taken at the node), both of these conditions are fulfilled by the presence of a gap and a kink.  In pumped data, the gap is partially or completely suppressed, which will partially or completely suppress the \PDH{}.

\begin{figure*}\centering\includegraphics[width=7in]{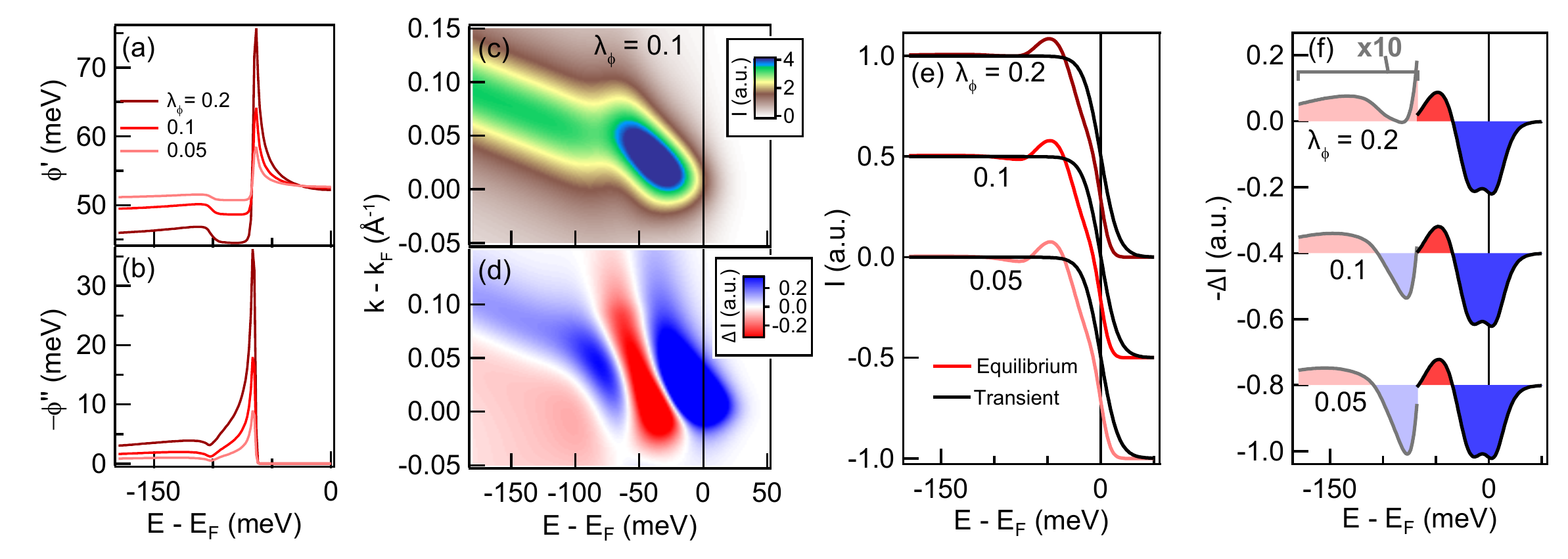}
\caption{Simulations of ARPES intensity using Eliashberg equations, an Einstein boson at 35 meV with coupling strength $\lambda_Z = 1$, and various values of $\lambda_\phi$.  (a) and (b) The real and imaginary parts of $\phi$ at 20 K.  A constant is added to $\phi'$ in order to fix the superconducting gap is 30 meV.  (c) The ARPES intensity map for $\lambda_\phi=0.1$.  (d) The pump-induced change in ARPES intensity.  The transient spectra are simulated by taking $\phi$ = 0 and $T$ = 80 K.  (e) The momentum integrated intensity in the equilibrium and transient state, calculated from Eq. (\ref{DIeq}).  (f) The pump-induced change in momentum integrated intensity.  Note that parts of the curves are magnified by a factor of 10.
}
\label{Fig4}
\end{figure*}

In order to make more specific predictions, we constructed the simulations shown in Fig. \ref{Fig4}.  The simulation uses $Z(\omega)$ and $\phi(\omega)$ calculated self-consistently from the Eliashberg equations \cite{Scalapino1969}.  The equilibrium simulations all use $T$ = 20 K and a gap size of 30 meV, while the transient simulations all use $T$ = 80 K and set $\phi$ to zero. We use an Einstein boson mode with 35 meV energy, which is about the energy of the bosonic mode thought to be responsible for the off-nodal kink, whether it is a magnetic resonance mode \cite{Fong1999} or phonon mode \cite{Devereaux2004}.  We also assume a momentum-independent gap size, which is equivalent to the limiting case where the antinodes dominate electron-boson scattering.  Following Ref. \cite{Sandvik2004}, we use separate electron-boson coupling constants to calculate $Z$ and $\phi$, with $\lambda_Z = 1$ and various values of $\lambda_\phi$.  The physical meaning of $\lambda_Z$ is the isotropic component of the electron-boson coupling, while $\lambda_\phi$ is the component with $d_{x^2 - y^2}$ symmetry.  $\lambda_\phi$ is, in essence, the extent to which this particular boson contributes to $d$-wave superconductivity.  In equilibrium simulations, we keep the superconducting gap size fixed by adding a real constant to $\phi$; this real constant may be interpreted as the contribution to superconductivity from other sources.  The real and imaginary components of $\phi$ at equilibrium are shown in Figs. \ref{Fig4}(a) and \ref{Fig4}(b), respectively.

Figure \ref{Fig4}(c) shows the simulated equilibrium ARPES intensity map for $\lambda_\phi=0.1$, and Fig. \ref{Fig4}(d) shows a map of the difference between equilibrium and transient intensity.  In order to make the simulation more realistic, we include a 35 meV contribution to the imaginary self-energy from impurity scattering, and a 15 meV instrumental energy resolution. Identical to the experimental procedure, we apply a normalization factor to each map so that they have the same average intensity between 200 and 300 meV.  The simulations suggest that it is inappropriate to normalize intensity at high binding energy; however, it was experimentally necessary to correct for laser power fluctuations.  The resulting intensity maps are remarkably similar to those observed in Fig. \ref{Fig1}.

Figure \ref{Fig4}(e) shows the momentum-integrated intensity in both the equilibrium and transient states.  In the equilibrium state, we see a \PDH{} similar to that observed in experiment, and in the transient state, the \PDH{} disappears because of the suppression of the superconducting gap.  The differences between the equilibrium and transient curves are shown in Fig. \ref{Fig4}(f), and they resemble the experimental observations in Fig. \ref{Fig2}.  If we identify $p_2$ as we did in Fig. \ref{Fig3}, we find that it appears around 80 meV, which is somewhat larger than the value of $\Delta+\Omega$ = 65 meV.  This may explain why in experimental measurements, $p_2$ does not precisely match the kink energy.

The most shocking prediction is that as $\lambda_\phi$ increases, the dip strength decreases.  By looking at the experimentally observed dip strength and comparing to simulations, we could in principle estimate the value of $\lambda_\phi$, which would tell us the extent to which the electron-boson kink contributes to high-temperature superconductivity.

\section{Discussion}

In this study, we explored the \PDH{} in cuprates by using two novel techniques. The first technique, TARPES, allows us to quickly compare measurements with and without a superconducting gap.  The second technique is to integrate the intensity over a single dimension of momentum, unlike previous studies which either took measurements at single momenta \cite{Dessau1991, Hwu1991, Dessau1992, Fedorov1999, Norman1997, Campuzano1999, Kaminski2001}, or used techniques that integrate over large regions of momentum \cite{Fischer2007, Loret2016}.  One-dimensional integration is a fruitful way to look at electron-boson coupling because it improves statistics, and because the intensity is expected to follow the simple expression in Eq. (\ref{DIeq}).

Using these techniques, we observe a pump-induced shift of intensity into the dip of the \PDH{}, similar to what is seen in equilibrium temperature-dependent measurements at the antinode \cite{Dessau1991, Hwu1991, Dessau1992, Fedorov1999, Norman1997, Campuzano1999, Kaminski2001}.  The maximum intensity change occurs near 70 meV for a wide range of momenta, in both underdoped and overdoped samples.  The presence over a wide doping range suggests that the observed structure is related to electron-boson coupling, and not to bilayer splitting or the pseudogap.  Based on simulations and considerations of Eq. (\ref{DIeq}), we find that the dip strength depends on the superconducting gap size, and on the parameter $\lambda_\phi$, which represents the degree to which electron-boson coupling contributes to $d$-wave superconductivity.

We also observe a similar change of intensity near 140 meV.  This may arise from a dip produced by a second bosonic mode \cite{He2013}, but if so it cannot be a phonon, given that the phononic modes reach no higher than 80 meV \cite{Renker1989}.  Another possibility is that the dip at 140 meV is a second-order effect of the same bosonic mode that produced the dip at 70 meV.  The first-order effect would be expected to appear near $\Delta+\Omega$, and the second-order effect is a replica near $\Delta+2\Omega$, although simulations and observations suggest these energies are not precise.  Our simulations do show second-order features, visible at 100 meV in Figs. \ref{Fig4}(a) and \ref{Fig4}(b), but to properly calculate second-order effects, it may be necessary to use polaronic simulations \cite{Verdi2017}.

The most exciting implication of these results, is that it may be possible to use the dip strength to determine how much electron-boson coupling contributes to $d$-wave superconductivity.  In fact, Ref. \cite{Bok2016} has already demonstrated the capability to use ARPES experiments to measure the bosonic contribution to superconductivity, but the method shown here is simpler and does not require as high resolution.  We caution against drawing specific conclusions from the present simulations, which make use of several simplifications.  First, we use a single constant gap size, when in the real system, the gap size is a function of momentum.  Second, in using Eq. (\ref{DIeq}), we assume the integration window is very large.  In practice, the integration window is finite, which could lead to a \PDH{} even in the nodal equilibrium data, which is experimentally observed in Fig. \ref{Fig2}.  Third, we do not include the self-energy contribution from electron-electron interactions, nor do we attempt to account for the pseudogap.  Nonetheless, with more testing and refinement of the calculations, we are hopeful that studies of the dip could address the longstanding question of how the electron-boson coupling relates to high-temperature superconductivity.

\begin{acknowledgments}
We thank A. Kemper for helpful discussion in the early stages of this work.  This work was supported by the U.S. Department of Energy, Office of Science, Basic Energy Sciences, Materials Sciences and Engineering Division under Contract No. DE-AC02-05-CH11231 within the Ultrafast Materials Science Program (KC2203).
\end{acknowledgments}

\bibliographystyle{apsrev4-1}
\bibliography{C:/Users/Tristan/Desktop/Dissertation/Thesis/tlmbib}

\end{document}